\documentclass[sigconf]{acmart}

\usepackage{amsmath}
\usepackage{amsfonts}

\usepackage{dsfont} 

\usepackage{mathtools} 

\usepackage{enumitem}

\usepackage{amsthm}

\usepackage{hyperref}

\usepackage{multirow}
\usepackage{booktabs}
\usepackage{graphicx}
\usepackage{caption}
\usepackage{subcaption}
\usepackage[outdir=./]{epstopdf}
\usepackage{siunitx}
\usepackage[ruled,vlined]{algorithm2e}
\usepackage{colortbl}
\usepackage{tabulary}
\usepackage{etoolbox}

\usepackage{booktabs,tabularx}

\usepackage[most]{tcolorbox}
\tcbset{
  colback=gray!5, colframe=gray!35, boxrule=0.3pt,
  left=6pt, right=6pt, top=4pt, bottom=4pt
}
\newtcolorbox{webconfcallout}[1][]{
  float,               
  floatplacement=t,    
  before skip=6pt, after skip=6pt,
  sharp corners,
  enhanced,
  #1
}

\usepackage{placeins} 
\usepackage{dblfloatfix} 

\copyrightyear{2026}
\acmYear{2026}
\setcopyright{cc}
\setcctype{by}
\acmConference[WWW '26] {Proceedings of the ACM Web Conference 2026}{April 13--17, 2026}{Dubai, United Arab Emirates.}
\acmBooktitle{Proceedings of the ACM Web Conference 2026 (WWW '26), April 13--17, 2026, Dubai, United Arab Emirates}
\acmISBN{}
\acmDOI{}
\begin{document}

\title{Dynamics of Human-AI Collective Knowledge on the Web: A Scalable Model and Insights for Sustainable Growth}

\author{Buddhika Nettasinghe}
\email{buddhika-nettasinghe@uiowa.edu}
\orcid{0000-0002-6070-892X}
\affiliation{%
  \institution{University of Iowa}
  \city{Iowa City}
  \state{Iowa}
  \country{USA}
}

\author{Kang Zhao}
\email{kang-zhao@uiowa.edu}
\orcid{0000-0002-8321-2804}
\affiliation{%
  \institution{University of Iowa}
  \city{Iowa City}
  \state{Iowa}
  \country{USA}
}

\renewcommand{\shortauthors}{Nettasinghe et al.}

\begin{abstract}
Humans and large language models (LLMs) now co-produce and co-consume the web’s shared knowledge archives. Such human-AI collective knowledge ecosystems contain feedback loops with both benefits~(e.g.,~faster growth, easier learning) and systemic risks~(e.g.,~quality dilution, skill reduction, model collapse). To understand such phenomena, we propose a minimal, interpretable dynamical model of the co-evolution of archive size, archive quality, model~(LLM) skill, aggregate human skill, and query volume. The model captures two content inflows (human, LLM) controlled by a gate on LLM-content admissions, two learning pathways for humans (archive study vs.\ LLM assistance), and two LLM-training modalities~(corpus-driven scaling vs. learning from human feedback). Through numerical experiments, we identify different growth regimes (e.g., healthy growth, inverted flow, inverted learning, oscillations), and show how platform and policy levers (gate strictness, LLM training, human learning pathways) shift the system across regime boundaries. Two domain configurations (PubMed, GitHub \& Copilot) illustrate contrasting steady states under different growth rates and moderation norms. We also fit the model to Wikipedia’s knowledge flow during pre-ChatGPT and post-ChatGPT eras separately. We find a rise in LLM additions with a concurrent decline in human inflow, consistent with a regime identified by the model. Our model and analysis yield actionable insights
for sustainable growth of human-AI collective knowledge on the Web.
\end{abstract}




\maketitle
\begin{tcolorbox}[colback=gray!10,colframe=gray!55,boxrule=0.5pt,left=1pt,right=1pt,top=1pt,bottom=1pt] \itshape Humans and AI now co-write and co-read web's knowledge archives (e.g., Wikipedia). Through a dynamical systems lens, we analyze such coupled human–AI knowledge systems and propose strategies for sustainable growth in both size and quality. \end{tcolorbox}
\section{Introduction}

\vspace{0.0cm}
The rapid proliferation of Large Language Models (LLMs) and other generative AI systems trained on vast online archives is reshaping how humans produce, consume, and share knowledge~\cite{burton2024large}. Each day, millions of individuals query LLMs about topics ranging from health and politics to coding and scientific research. The outputs of these interactions are increasingly fed back into the same online archive, forming a shared knowledge space that both humans and AI models learn from and contribute to. For example, a researcher might use LLMs to refine their ideas, and resulting publications may then be used for future research as well as to train future LLMs. Though this tight coupling between human and AI knowledge might accelerate the growth of their shared online archive, it also introduces systemic vulnerabilities~\cite{hacker2025ai}. One well-documented risk is \emph{model collapse}, the degradation of generative AI performance when models are trained on data containing the outputs from previous models~\cite{shumailov2024ai}. However, beyond this AI-centric phenomenon lie deeper society-scale concerns. Two particularly pressing risks are:
\vspace{-0.05cm}
\begin{enumerate}
    \item \emph{Quality dilution}: unvetted LLM outputs may increasingly dominate the shared archive, eroding the ratio of high-fidelity human scholarship to synthetic content.\\ e.g.,~the overall content quality on collective knowledge archives such as Wikipedia, PubMed and online code repositories such as GitHub might decline or homogenize if they are primarily generated by LLMs.

    \item \emph{Human competence inversion}: human expertise may degrade as a consequence of learning from a low-quality archive, potentially reducing the quality of their future contributions to the same archive, triggering a feedback loop of declining expertise and knowledge quality.\\ e.g.,~researchers having to rely on pre-print or code repositories that are being heavily diluted by LLM generated content.
\end{enumerate}
\vspace{-0.0cm}
\begin{figure*}[htbp]        
  \centering                
\includegraphics[width=1\linewidth, trim = {0.2cm 3.6cm 0cm 0.1cm}, clip ]{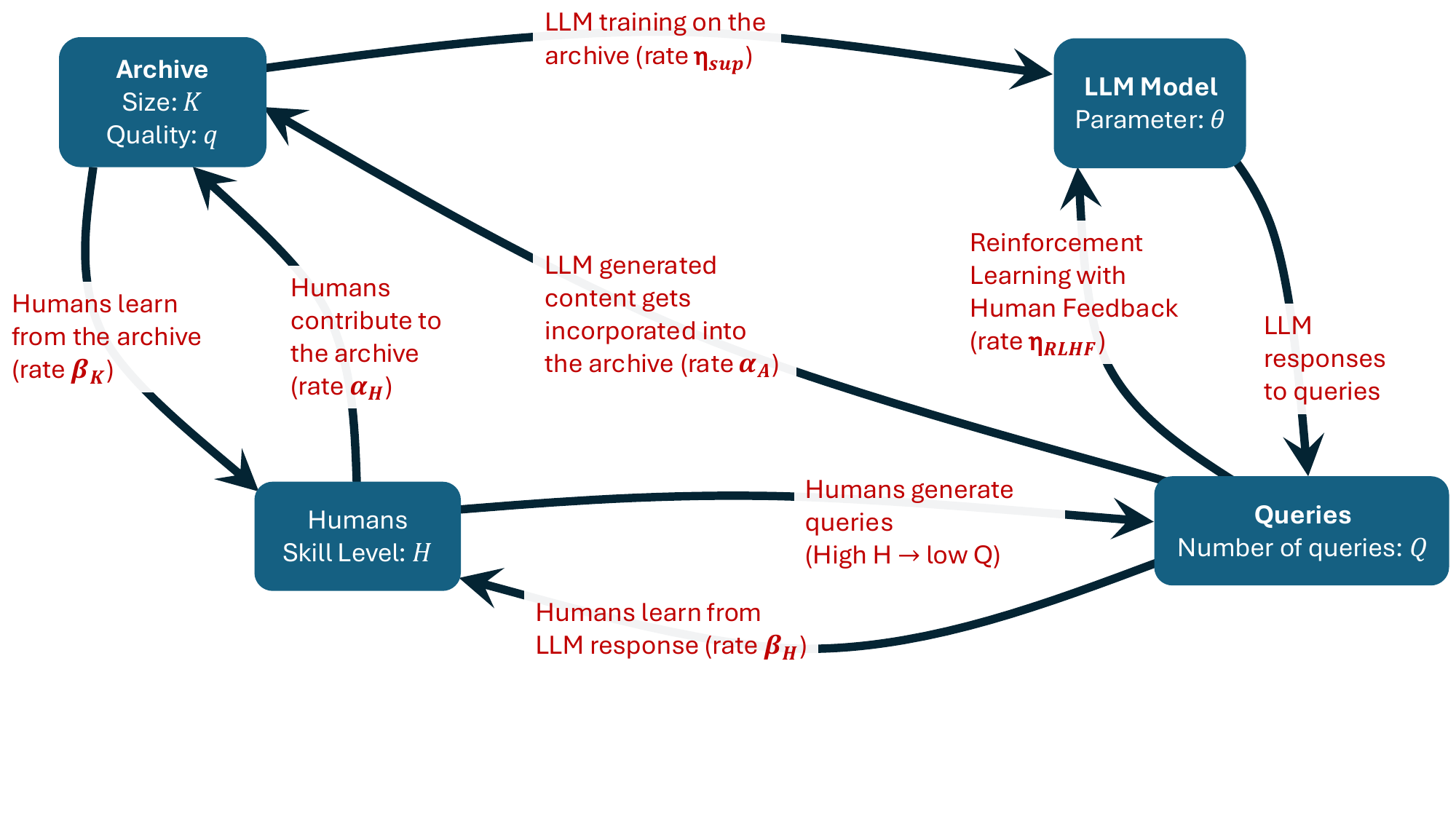} 
  \caption{Overview of the proposed model of human-AI collective knowledge (Eq.~\eqref{eq:model}). Different parameter configurations of the model produce distinct, interpretable growth regimes in the human-AI collective knowledge ecosystem. The model provides a unified view of phenomena such as model collapse, quality dilution and competence inversion, and yields insights on strategies for sustainable growth. It can be calibrated using data to understand real-world platform dynamics as discussed in Sec.~\ref{sec:results}.}
  \label{fig:schematic}  
  \vspace{-0.cm}
\end{figure*}
Emergence of such systemic risks and their collective dynamics (instead of each risk in isolation) are not well captured by existing studies~\cite{weidinger2021ethical}. Most prior works focus either on the LLMs (e.g.,~model collapse~\cite{shumailov2024ai}, scaling laws of machine learning models~\cite{kaplan2020scaling}) or the human learning (e.g.,~cognitive load theory~\cite{sweller2011cognitive}) in isolation. However, a framework that views humans and LLMs as operating on
the same collectively shared archive has been absent in literature. Such a framework is critically needed to obtain a unified view of various systemic risks and to preserve the healthy growth of both quantity and quality of human-AI  collective knowledge.

\vspace{0.1cm}
\noindent
{\bf Main Contributions: }\\  
1.) \emph{A dynamical model of human-AI collective knowledge: } We introduce a minimal five-variable dynamical system that captures the core feedback loops between human skill $H$, LLM model skill~$\theta$, archive size $K$, archive quality $q$, and demand for LLM information~$Q$. A graphical illustration of how these variables depend on each other is shown in Fig.~\ref{fig:schematic}. The proposed model draws on insights from both machine learning (e.g., scaling laws of machine learning models) and cognitive psychology (e.g., human learning curves and saturation) to characterize the dynamics of a human-AI collective knowledge ecosystems under various scenarios. 

The proposed model provides a unified view of systemic risks, including model collapse~\cite{shumailov2024ai}, human competence inversion and quality dilution, via a single holistic framework. For example, 
model collapse is a consequence of the upper loop in Fig.~\ref{fig:schematic}: low-fidelity LLM outputs admitted to the archive reduce its quality $q$, producing models trained on it with lower skill $\theta$; their outputs, propagated via the query flow $Q$, further erode the archive. 
Human competence inversion is the reduction of human skill $H$ as a consequence of humans depending heavily on either the low-quality (smaller $q$) archive or a weakened model (smaller $\theta$)~i.e.,~the two arrows that point towards human skill $H$ in Fig.~\ref{fig:schematic}. The degraded human skill then would further reduce the quality of human contributions to the archive, causing the model collapse and human competence inversion both to amplify. Hence, model collapse, quality dilution and human competence inversion are all related to each other, and as such, strategies to battle them should not view them in isolation. 

\vspace{0.1cm}
\noindent
2.)~\emph{Insights for healthy growth and preventing systemic risks: }Using the proposed model, we identify critical regimes—such as healthy growth, stagnation, rebound, oscillations, collapse—in the human-AI collective knowledge, and offer insights on regulatory-level and platform-level interventions that can safeguard its growth in terms of both quality and quantity.

\vspace{0.1cm}
\noindent
3.)~\emph{Calibration of the model with real-data: }By calibrating the model using data from Wikipedia, we show that the LLM contributions have increased in the post-ChatGPT era compared to pre-ChatGPT era. In addition, we also provide two real-world case studies based on PubMed and GitHub Copilot, showing that they belong to two distinct growth regimes captured by the proposed model.

\vspace{0.15cm}
\noindent
{\bf Motivation from the Literature: }The critical need for dynamical system models of the human-AI collective knowledge has been emphasized in recent literature~\cite{barfuss2025collective}. Dynamical system models are particularly suitable for understanding the consequences of feedback loops in complex systems, such as those that have been shown to exist in the human-AI collective knowledge ecosystems~\cite{pedreschi2024human, glickman2025human}. Recent work has attempted to model some aspects of human-AI interaction using dynamical systems~e.g.,~prompt adaptation~\cite{jahani2024prompt}, search and information retrieval behaviors~\cite{garetto2025information}. 

However, model collapse and related phenomena have largely been studied only in terms of the properties of the underlying probability distributions~\cite{shumailov2024ai}. Dynamical systems have been underutilized to understand and evaluate collective human-AI knowledge. The successful uses of dynamical systems to understand societal problems such as inequality~\cite{nettasinghe2021emergence,avin2015homophily}, polarization~\cite{yang2020us, nettasinghe2025out, nettasinghe2025group}, information diffusion~\cite{lopez2008diffusion} and social learning dynamics~\cite{lu2025first, yang2021dynamical} further motivate our approach for understanding human-AI collective knowledge.

\section{A Dynamical Model of Human-AI Collective Knowledge}

This section formally presents the model of human-AI collective knowledge~(illustrated in Fig.~\ref{fig:schematic}), followed by a detailed discussion of its key variables, assumptions, and practical applications. We also briefly discuss how the model can be generalized to more scenarios.

\vspace{-0.1cm}
\subsection{Model}
\label{subsec:model}
We propose a scalable dynamical model that captures how the archive size $K(t)$, archive quality $q(t)$, LLM model capacity~$\theta(t)$, aggregate human skill $H(t)$, and LLM query volume~(demand) $Q(t)$ co-evolve on a common time axis $t$ (e.g.,~months). We use a dot to denote the time derivative~(e.g.,~\ $\dot{K}(t)=\tfrac{\mathrm{d}K}{\mathrm{d}t}$). Eq.~\eqref{eq:model} formalizes the five–state dynamical system,
\begin{figure*}[t]
\centering
\Large
\vspace{-0.2cm}
\begin{minipage}{\textwidth}
\begin{subequations}\label{eq:model}
\begin{align}
\dot{K}(t) &=
\underbrace{\alpha_{H}H(t)}_{\substack{\text{Human contributions}\\\text{More skill $\Rightarrow$ more content}}}
\;+\;
\underbrace{\alpha_{A}\,Q(t)\,g\,\!\bigl(a(t)\bigr)}_{\substack{\text{LLM contributions}\\\text{More queries \& lax gate $\Rightarrow$ more AI content}}}
\;-\;\delta_{K}\,K(t),
& \text{\textbf{Archive size $K(t)$}} \label{eq:model_K}\\[6pt]
\dot{q}(t) &=
\frac{\alpha_{H}H(t)}{K(t)}
\underbrace{\bigl(q_{H}-q(t)\bigr)}_{\substack{\text{Human content pulls}\\\text{quality toward }q_H}}
\;+\;
\frac{\alpha_{A}\,Q(t)\,g\,\!\bigl(a(t)\bigr)}{K(t)}
\underbrace{\bigl(a(t)-q(t)\bigr)}_{\substack{\text{AI content pulls}\\\text{quality toward }a(t)\\\text{(better model $\Rightarrow$ higher $q$)}}}
\;-\;\delta_{q}\,q(t),
& \text{\textbf{Archive quality $q(t)$}} \label{eq:model_q}\\[6pt]
\dot{\theta}(t) &=
\underbrace{\eta_{\text{sup}}\bigl(\theta^{*}(K(t),q(t))-\theta(t)\bigr)}_{\substack{\text{LLM training on archive moves }\theta\text{ toward }\theta^{*}(K,q).\\\text{Scaling laws govern how }\theta^{*}(K,q)\text{ grows with }K,q.}}
\;+\;
\underbrace{\eta_{\text{RLHF}}\mathrm{RLHF}\!\bigl(\theta(t),Q(t)\bigr)}_{\substack{\text{Interaction-driven gain (RLHF)}\\\text{More queries $\Rightarrow$ faster approach to frontier}}}
& \text{\textbf{LLM capacity $\theta(t)$}} \label{eq:model_theta}\\[6pt]
\dot{H}(t) &=
\underbrace{\beta_{K}\bigl(H_{\max}(K(t),q(t))-H(t)\bigr)}_{\substack{\text{Studying the archive moves}\\\text{human skill toward }H_{\max}(K,q)}.}
\;+\;
\underbrace{\beta_{A}\,a(t)\,S(Q(t))}_{\substack{\text{Learning assistance from LLMs}\\\text{Gains scale with accuracy and queries.}}}
\;-\;\gamma_{H}\,H(t),
& \text{\textbf{Human skill $H(t)$}} \label{eq:model_H}\\[6pt]
\dot{Q}(t) &=
\underbrace{\xi\,\!\bigl(H(t)\bigr)}_{\substack{\text{Baseline query inflow}\\\text{More skill $\Rightarrow$ fewer queries}}}
\;-\;\rho_{Q}\,Q(t),
& \text{\textbf{Query volume $Q(t)$}} \label{eq:model_Q}
\end{align}
\end{subequations}
\end{minipage}
\end{figure*}
where the auxiliary functions are defined in Table~\ref{tab:model_functions}.
\begin{table*}[!htbp]
\centering
\renewcommand{\arraystretch}{1.2} 
\caption{Key functional components of the human–AI collective knowledge-dynamics model given in Eq.~\eqref{eq:model}. Each component is motivated by the literature as discussed in Sec.~\ref{subsec:variable_discussion} and Sec.~\ref{subsec:model_discussion}. The modular nature of the functional components also makes the model generalizable~e.g., the form of $\theta^{*}(K,q)$ can be changed to explore a different scaling law for LLMs. A detailed list of terminology is given in Table~\ref{tab:terminology} in Appendix~\ref{app:terminology}.}
\vspace{-.25cm}
\begin{tabularx}{\textwidth}{@{} l l X @{}}
\toprule 
\textbf{Function} & \textbf{Formula} & \textbf{Description \& Rationale (more details in Sec.~\ref{subsec:variable_discussion})} \\
\midrule
LLM skill curve &
$\displaystyle \sigma(\theta)=\frac{1}{1 + e^{-(\theta - \theta_{\text{mid}})}}$ &
Maps the latent (raw) model capacity $\theta$ to base correctness with an S-curve with midpoint (inflection point) $\theta_{\text{mid}}$. \\

LLM answer accuracy &
$a(\theta,q)=\sigma(\theta)\,q$ &
 Overall answer accuracy as the product of model skill and archive quality, capturing the dependence of LLM answer accuracy on both model skill and quality of the content retrieved from the archive (RAG)~\cite{lewis2020retrieval}. \\

Admission gate &
$g(a)=\dfrac{1}{1+\exp[-\kappa_{\text{gate}}(a-a_{0})]}$ &
Share of AI outputs admitted to the archive (by platform moderators, etc.) as a logistic function of accuracy; higher $a_0$ means stricter admission thresholds and larger $\kappa_{\text{gate}}$ means sharper enforcement. \\

Scaling-law target &
$\displaystyle \theta^{*}(K,q)=\theta_{\max}\,
      \frac{\ln(1+K)}{\ln(1+K_{\max})}\,q$ &
 Larger, higher-quality corpora raise attainable training skill with diminishing returns in $K$, consistent with empirical scaling laws~\cite{kaplan2020scaling,hestness2017deep} that govern model training. \\

RLHF gain &
$\displaystyle \mathrm{RLHF}(\theta,Q)=
      \frac{Q}{Q+Q_{1/2}}\,
      \bigl(\theta_{\max}-\theta\bigr)$ &
Human feedback through queries accelerates model learning when it lags far from the frontier; the benefit saturates once $Q$ exceeds half-saturation~$Q_{1/2}$~\cite{don2025future}. \\

Human ceiling & $H_{\max}(K,q)=H_{\infty}\,\frac{(qK)^{\beta}}{K_{1/2}^{\beta}+(qK)^{\beta}}$ &
The volume ($K$) and clarity ($q$) of the archive jointly raise maximum achievable skill with diminishing returns in the form of a Hill (generalized Michaelis–Menten) function that reproduces human learning curves~\cite{lane2012skill}.\\

Learning from AI & $S(Q) = \dfrac{Q}{Q+Q_{\text{sat}}}$ & LLM assistance yields a tutoring boost to human learning that rises with query volume~$Q$ but saturates beyond a half-saturation level $Q_{\text{sat}}$, reflecting that LLM additional help has declining marginal benefits. \\

Baseline query rate &
$\xi(H)=\xi_{0}\bigl(1+T_\text{difficulty}\bigr)\,e^{-\kappa_{H}H}$ &
More skilled users need fewer LLM queries; harder tasks raise the baseline. \\
\bottomrule
\end{tabularx}
\label{tab:model_functions}
\end{table*}
The model is well-posed and invariant as discussed in Appendix~\ref{app:wellposedness}.


\subsection{Discussion of the Model, Key Variables and Auxiliary Functions}
\label{subsec:variable_discussion}

The variables of the model in Eq.~\eqref{eq:model} and their dynamics are intuitive and motivated from literature as discussed below.

\vspace{0.2cm}
\noindent
 \textbf{Archive size ${K(t)}$} is the total number of unique, citable knowledge items
(e.g.,~tokens, research papers, code repositories, or wiki pages)
that remain accessible in the online archive at time~$t$. The dynamics of $K(t)$ are given by Eq.~\eqref{eq:model_K}. New items enter the archive through (i)~\emph{human} contributions at rate $\alpha_H H(t)$, or, (ii)~\emph{LLM} content admitted through the archive quality gate $g\!\bigl(a(t)\bigr)$ at rate $\alpha_A Q(t) g\!\bigl(a(t)\bigr)$.

Thus, higher human skill boosts human content,
whereas a large query volume coupled with a lenient gate amplifies AI-generated additions.
The gate $g(a)$ represents platform moderation mechanisms (e.g.,~editors, peer reviewers, community moderators, etc.).
Items exit the archive, for example via retractions, link rot, or obsolescence of knowledge, at rate $\delta_K K(t)$ (third term of Eq.~\eqref{eq:model_K}).
The coefficients $\alpha_H$ and $\alpha_A$ scale the relative contribution rates of humans
and AI e.g., $\alpha_H \ll \alpha_A$ would indicate AI content additions outpacing human contributions.
 
\vspace{0.2cm}
\noindent
 \textbf{Archive quality $q(t)\!\in\![0,1]$} measures the mean fidelity of the
knowledge items being added at time~$t$.
It is an inertial average that drifts toward two reference levels:  
(i)~the intrinsic benchmark of human-only contributions, $q_H$, and  
(ii)~the current LLM answer accuracy, $a(t)=\sigma\!\bigl(\theta(t)\bigr)\,q(t)$.
Whenever humans add material, the first term of Eq.~\eqref{eq:model_q} nudges the running average upward (or downward) toward~$q_H$. The magnitude of this nudge is proportional to their productivity $\alpha_H H(t)$, and inversely proportional  
to the archive size $K(t)$, reflecting the fact that archive quality is difficult to be changed when the archive is large. 
Analogously, AI additions pull quality toward the model’s current accuracy $a(t)$ via the
second term of Eq.~\eqref{eq:model_q}, which is amplified by a large query volume $Q(t)$ and a permissive gate
$g(a)$. For example, a high query load (larger $Q(t)$) on a poor model whose accuracy is low compared to current archive quality ($a(t)< q(t)$) can reduce the quality of the archive if the gate lets the most of the poor quality content in. Finally, existing items become obsolete or are re-evaluated at rate
$\delta_q q(t)$ (third term of Eq.~\eqref{eq:model_q}),
capturing the gradual quality decay due to time.
Thus $\alpha_H$ and $\alpha_A$ also regulate how forcefully new human
or AI content steers the quality trajectory, while $K(t)$ modulates their
marginal impact as the archive expands.

\vspace{0.2cm}
\noindent
 \textbf{LLM skill, $\theta(t)$} is an abstract measure of model capacity~
(e.g.,\ loss-equivalent parameter count or effective compute) that ultimately
controls baseline answer accuracy through the logistic “skill curve”
$\sigma\!\bigl(\theta\bigr)=\bigl[1+\exp\!\bigl(-(\theta-\theta_{\mathrm{mid}})\bigr)\bigr]^{-1}$. 
Its evolution is governed by Eq.~\eqref{eq:model_theta}, whose two terms
represent distinct training modes: \\
\textit{(i)~Supervised pre-training on the archive} (first term in Eq.~\eqref{eq:model_theta}) pulls the model skill $\theta(t)$ towards the scaling-law
target $\theta^{*}(K,q)$, which rises linearly with archive quality $q$ and logarithmically with archive size $K$ and saturates as $K$ approaches $K_{\max}$, mimicking the scaling-laws of deep learning based natural language models~\cite{kaplan2020scaling, hestness2017deep}. In particular, the scaling-law emphasizes the diminishing returns of the size of the training dataset and the importance of quality.\\ \textit{(ii)~Reinforcement Learning with Human Feedback~(RLHF - second term in Eq.~\eqref{eq:model_theta})} accelerates progress via human interactions:
the feedback gain
$\mathrm{RLHF}(\theta,Q)=\dfrac{Q(t)}{Q(t)+Q_{1/2}}\bigl(\theta_{\max}-\theta(t)\bigr)$
couples user demand (query load $Q(t)$) to the distance from the
current frontier $\theta_{\max}$, with half-saturation constant $Q_{1/2}$. A higher demand for LLM responses can thus improve the model via RLHF, although its effect slows down after $Q_{1/2}$~\cite{don2025future}.\\
The learning rates $\eta_{\text{sup}}$ and $\eta_{\text{RLHF}}$ set the
relative speeds of corpus-driven versus feedback-driven improvements, which is a key design lever important in LLMs.

\vspace{0.2cm}
\noindent
 \textbf{Average human skill $H(t)$} 
captures the collective human skill at
time~$t$. Its trajectory follows Eq.~\eqref{eq:model_H}, driven by three
competing processes:\\
\textit{(i)~Learning from the archive} (first term of Eq.~\eqref{eq:model_H})
corresponds to the classical self-study using the knowledge available in the archive~e.g.,~researchers relying on the previously published research. It pulls the human skill $H(t)$ towards a ceiling $H_{\max}(K,q)$ (see Table~\ref{tab:model_functions}) which rises with both archive size $K$ and fidelity $q$ and saturates at
$H_{\infty}$. 
In particular, $H_{\max}(K,q)$ is a Hill (generalised Michaelis–Menten) function which captures the well-documented diminishing returns of human learning curves (i.e.,~the plateauing of skill with rising quality and quantity of learning materials~\cite{lane2012skill}). Exponent $\beta$ modulates steepness of  the curve and $K_{1/2}$ is the half-saturation point. $H_{\max}(K,q)$ reduces to the classical Michaelis–Menten response function when $\beta = 1$.  \\
\textit{(ii)~Learning from LLM answers} (second term of
Eq.~\eqref{eq:model_H}) adds a stimulus that scales with LLM answer accuracy $a(t)$ and with the
query volume $Q(t)$ up to a saturation threshold $Q_{\text{sat}}$,
mirroring cognitive-load limits on how many AI explanations users can absorb.

Forgetting and skill turnover are captured by the decay term
$-\gamma_H H(t)$, representing memory loss, skill obsolescence, and
user churn. Hence $\beta_K$ and $\beta_A$ in Eq.~\eqref{eq:model_H} set the relative influence of archival study
versus reliance on LLMs, and $\gamma_H$
regulates how quickly human skill erodes when new information inflow slows.

\vspace{0.2cm}
\noindent
 \textbf{Query intensity $Q(t)$} given in Eq.~\eqref{eq:model_Q} is the smoothed usage intensity of LLMs at time $t$. New queries are initiated at rate $\xi(H)=\xi_{0}\bigl(1+T_\text{difficulty}\bigr)\,e^{-\kappa_{H}H}$, reflecting the fact that more knowledgeable users rely less on LLMs. The decay rate~$\rho_{Q}$ corresponds to how quickly the query intensity decays in the absence of new demand.
 
\subsection{Practical Applications and Generalizations}
\label{subsec:model_discussion}


 The dynamics of the five state variables specified by Eq.~\eqref{eq:model}~(also illustrated in Fig. \ref{fig:schematic}) yields an interpretable model of a human-AI collective knowledge ecosystem. In practice, it can be used in several ways as we discuss next.

\vspace{0.1cm}
\noindent
{\bf Design and Policy Levers: } The dynamical system in Sec. \ref{subsec:model}  is intentionally parsimonious: three pairs of control parameters govern its qualitative behavior.
Relative content flow $(\alpha_H,\alpha_A)$ sets the balance of human vs.\ AI contributions to the archive; learning pathways $(\beta_K,\beta_A)$ determine whether users rely more on self-study or on LLM explanations for learning; training modality weights $(\eta_{\text{sup}},\eta_{\text{RLHF}})$ allocate model improvement between offline corpus training and interaction-driven feedback.  Each pair corresponds to a real-world design lever (e.g.,~user engagement patterns, and RLHF investment), allowing policy and design experiments. Our subsequent results show how the right calibration of these parameters can ensure healthy growth of the shared archive as well as how the model can be calibrated using data from real-world platforms. 

\vspace{0.1cm}
\noindent
{\bf Modularity: }All remaining parameters appear only inside the auxiliary functions listed in Table \ref{tab:model_functions}. Since these auxiliary functions are modular, one can easily swap in alternative forms to probe “what-if” scenarios~e.g., replacing the current LLM scaling law with a \emph{logarithmic} dependence of model skill on training-data quality, or substituting a non-Hill learning curve for human expertise.

\vspace{0.1cm}
\noindent
{\bf Extensions: }The model can be decomposed further to incorporate more nuances~e.g.,~\\
\textit{Multiple knowledge fields.} A vector state $K=(K_{\text{math}},K_{\text{CS}})$ with cross‑quality coupling allows interacting quality and quantity of fields. For example, a slowly evolving mathematics archive with few LLM contributions may coexist with a rapidly growing computer science archive; each field’s quality feeds back into the other via coefficients in the quality equation.\\
\textit{Experts versus novices.} Partitioning human skill $H(t)$ into experts ($E(t)$) and novices ($N(t)$) lets learning regime~$\beta_K, \beta_A$ vary by cohort: novices rely heavily on LLM explanations whereas experts rely more on primary literature. Further, the contribution quality (i.e.,~$q_H$) can be decomposed likewise in to two components $q_E, q_N$, with $q_E>q_N$ to denote the high quality of expert contributions compared to novices.


\vspace{-0.1cm}
\section{Numerical and Empirical Results}
\label{sec:results}

In this section, we first identify different growth regimes and their corresponding parameter regions, yielding insights on platform and policy designs to sustain growth~(Sec.~\ref{subsec:numerical_experiments}). We then provide case studies based on two real-world platforms (PubMed and GitHub \& Copilot: Sec.~\ref{subsec:hybrid_results}). Finally, we calibrate the knowledge-flow of the model to the Wikipedia platform to compare its dynamics in the pre- and post- chatGPT eras~(Sec.~\ref{sec:empirical-wikipedia}).

\subsection{Numerical Results}
\label{subsec:numerical_experiments}

\begin{figure*}[]
  \centering
  \begin{subfigure}[b]{0.32\textwidth} 
    \centering
    \includegraphics[width=\textwidth,
                     trim={0.2cm 0.3cm 0.2cm 0.cm}, clip]{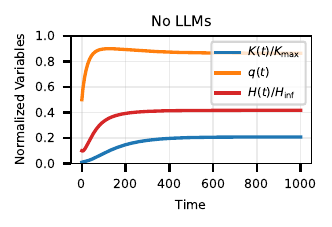}
    \vspace{-0.5cm}
    \caption{Pre-LLM Era: people learn from and contribute to the archive. Archive quality increases over time.}
    \label{subfig:No_LLMs}
  \end{subfigure}\hfill
    \begin{subfigure}[b]{0.32\textwidth} 
    \centering
    \includegraphics[width=\textwidth,
                     trim={0.2cm 0.3cm 0.2cm 0.cm}, clip]{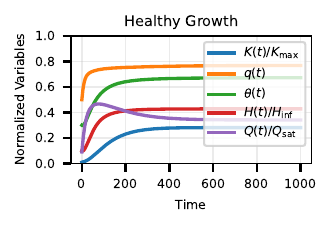}
    \vspace{-0.5cm}
    \caption{Archive size and human skill grow to be larger compared to case~(a) at the expense of a small quality decrease.}
    \label{subfig:Healthy Growth}
  \end{subfigure}\hfill
    \begin{subfigure}[b]{0.32\textwidth}  
    \centering
    \includegraphics[width=\textwidth,
                     trim={0.2cm 0.3cm 0.2cm 0.cm}, clip]{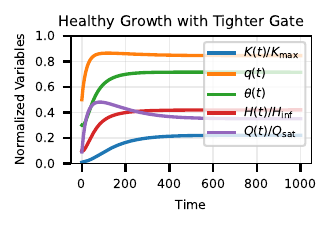}
    \vspace{-0.5cm}
    \caption{Archive quality and model skill are both higher compared to case (b) but the archive size is smaller due to lower admissions.}
    \label{subfig:HGTG}
  \end{subfigure}\hfill
    \begin{subfigure}[b]{0.32\textwidth}   
    \centering
    \includegraphics[width=\textwidth,
                     trim={0.2cm 0.3cm 0.2cm 0.cm}, clip]{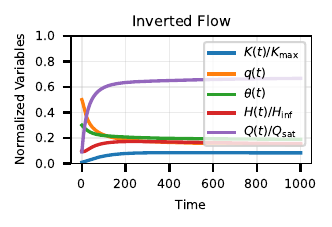}
    \vspace{-0.5cm}
    \caption{LLMs create more content compared to humans. Archive size slightly increases but its quality and model skill reduces. Human skill becomes stagnant. }
    \label{subfig:Inverted Flow}
  \end{subfigure}\hfill
    \begin{subfigure}[b]{0.32\textwidth}  
    \centering
    \includegraphics[width=\textwidth,
                     trim={0.2cm 0.3cm 0.2cm 0.cm}, clip]{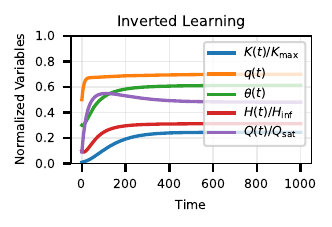}
    \vspace{-0.5cm}
    \caption{Humans learn more from LLMs than by studying the archive. Poor archive quality makes the skill stagnant and consequently the quality stagnant.}
    \label{subfig:Inverted_Learning}
  \end{subfigure}\hfill
    \begin{subfigure}[b]{0.32\textwidth}   
    \centering
    \includegraphics[width=\textwidth,
                     trim={0.2cm 0.3cm 0.2cm 0.cm}, clip]{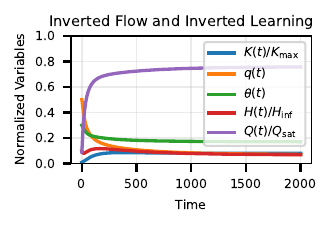}
    \vspace{-0.5cm}
    \caption{Worst-case scenario leading to the erosion of human skill, model skill and archive quality in a mutually reinforcing way. Query volume explodes due to eroding human skill. }
    \label{subfig:Inverted_Flow_and_Inverted_Learning}
  \end{subfigure}  \hfill
    \begin{subfigure}[b]{0.32\textwidth}  
    \centering
    \includegraphics[width=\textwidth,
                     trim={0.2cm 0.3cm 0.2cm 0.cm}, clip]{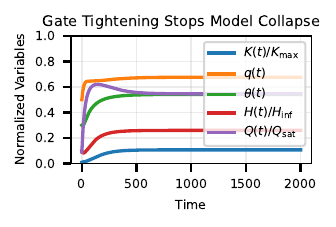}
                     \vspace{-0.5cm}
    \caption{Tightening the archive gate avoids the collapse of LLM skill, human skill and archive quality even with inverted flow and learning.}
    \label{subfig:Gate Tightening Stops Model Collapse}
  \end{subfigure}\hfill
    \begin{subfigure}[b]{0.32\textwidth}   
    \centering
    \includegraphics[width=\textwidth,
                     trim={0.2cm 0.3cm 0.2cm 0.cm}, clip]{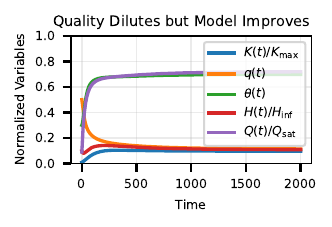}
    \vspace{-0.5cm}              \caption{High RLHF can keep the model skill increasing even when other quantities decrease or stagnate.}
    \label{subfig:Quality Dilutes but Model Improves}
  \end{subfigure} \hfill
      \begin{subfigure}[b]{0.32\textwidth} 
    \centering
    \includegraphics[width=\textwidth,
                     trim={0.2cm 0.3cm 0.2cm 0.cm}, clip]{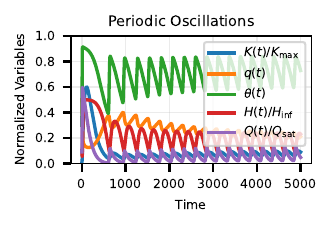}
    \vspace{-0.5cm}
    \caption{With substantially high RLHF, query oscillations drive recurrent archive-quality oscillations.}
    \label{subfig:Periodic Oscillations}
  \end{subfigure}
  \vspace{-0.25cm}
  \caption{Examples of different growth regimes in the human-AI collective knowledge ecosystem modeled by Eq.~\eqref{eq:model} (and illustrated in Fig.~\ref{fig:schematic}) for different parameter configurations. These results yield insights on policy- and platform- level interventions that can ensure sustainable growth of the size and quality of the shared online archive, human skill, and model capacity.}
  \label{fig:numerical_results}
    \vspace{-0.3cm}
\end{figure*}

We first simulate nine representative scenarios (see Fig.~\ref{fig:numerical_results}), each corresponding to a qualitatively distinct regime of human–AI knowledge co-evolution (Eq.~\eqref{eq:model} and Fig.~\ref{fig:schematic}). The normalized trajectories of key state variables—archive size $K(t)/K_{\max}$, archive quality $q(t)$, model skill $\theta(t)$, human skill $H(t)/H_{\infty}$, and query saturation $Q(t)/Q_\mathrm{sat}$—are compared across these settings to reveal how different structural configurations lead to healthy growth, stagnation, collapse, etc. The initial values are fixed across all nine configurations. A detailed explanation of the experimental setup with exact parameter configurations for each case is given in Appendix~\ref{app:numerical_simulation_details} for reproducibility.

\paragraph{(a) Pre-LLM Era:}
This baseline scenario excludes AI contributions ($\alpha_A = 0, \alpha_H = 0.5$) and disables AI learning ($\eta_{\mathrm{sup}} = \eta_{\mathrm{RLHF}} = 0$). Humans learn only by studying the archive~($\beta_A = 0, \beta_K = 0.05$). Thus, the archive growth and quality improvement are driven entirely by human expertise. Over time, both the normalized archive size and human skill gradually increase, and archive quality asymptotically approaches the intrinsic fidelity of human content ($q_H = 0.95$). This slow but steady regime represents the pre-LLM knowledge ecosystem.

\vspace{-0.2cm}
\paragraph{(b) Healthy Growth:}
Introducing modest LLM activity ($\alpha_A = 0.05$, $\eta_{\mathrm{sup}} = 0.05$, $\eta_{\mathrm{RLHF}} = 0.03$) and human reliance on LLMs to learn ($\beta_A = 0.03$) leads to accelerated growth in both human skill and archive size, albeit with a slight trade-off in archive quality (compared to pre-LLM era). This regime, where humans contribute more to the archive $\alpha_H\gg\alpha_A$ and they learn most from the archive  ($\beta_K>\beta_A$) leads to healthy growth: humans benefit from LLM explanations, while the LLM improves through RLHF and access to a growing, high-quality archive. The feedback loop between archive, humans and the LLM sustains a stable regime where the entire system co-evolves constructively. We treat this as our base case for comparison with the rest of the cases.

\vspace{-0.2cm}
\paragraph{(c) Healthy Growth with a Tighter Gate:} Tightening the gate (changing $a_0$ from 0.5 to 0.8) further increases the quality of the archive $q(t)$ (compared to (b)) by preventing less accurate content from ending up in the archive. The model skill $\theta(t)$ also increases further since it is trained on a better quality archive.  However, the archive size $K(t)$ settles at a smaller value since most LLM contributions do not pass the gate.

\vspace{-0.2cm}
\paragraph{(d) Inverted Flow:}
In this scenario, the archival flow is inverted: LLMs dominate content creation ($\alpha_H = 0.05< \alpha_A = 0.5$), while humans play a secondary role. As a result, the archive size increases only slightly (not enough content passes the archive gates) and quality $q(t)$ declines rapidly due to the admission of lower-accuracy LLM content. Since the model is trained on this degraded archive, its skill $\theta(t)$ declines as well. The human skill $H(t)$ stalls at suboptimal levels, as both the archive and model fail to provide sufficient boost in learning. This captures the onset of \emph{model collapse} and \emph{competence inversion} due to over-reliance on synthetic contributions. Early indications of inverted flow have already been seen in social media~\cite{moller2025impact} and academic writing~\cite{kobak2025delving}.

\vspace{-0.2cm}
\paragraph{(e) Inverted Learning:}
Here, users rely more heavily on LLM explanations than on primary archival study ($\beta_K = 0.03 < \beta_A = 0.05$), while content flow remains at the original healthy levels~($\alpha_H = 0.5 > \alpha_A = 0.03$). Human learning from LLMs does not achieve the skill level $H(t)$ in healthy growth.\footnote{Since LLM content accuracy $a(\theta,q)$ reaches inherent human quality $q_H = 0.95$ only for a well trained model relying on a high quality archive.} The stagnant human skill then leads to a self-perpetuating plateau in archive quality, which in turn makes model skill stagnant. This highlights the \emph{risks of over-reliance on AI} even when content quality is near optimal.

\vspace{-0.2cm}
\paragraph{(f) Inverted Flow and Inverted Learning (worst case scenario):}
Finally, combining both inverted flow ($\alpha_H = 0.05< \alpha_A = 0.5$) and inverted learning ($\beta_K = 0.03 < \beta_A = 0.05$) produces a catastrophic regime where all key metrics deteriorate. The archive is saturated with low-quality LLM outputs, humans rely more on the poor LLM explanations, and human skill decays due to both quality in LLM answers and archive content. As human skill falls, the volume of queries increases, but this only amplifies the injection of low-quality LLM content into archive, fueling the degradation. Model skill declines over time due to poor training data in the archive, leading to a \emph{self-reinforcing collapse in the human–AI collective knowledge ecosystem} via multiple vicious feedback loops.

\vspace{-0.2cm}
\paragraph{(g) Gate-keeping to sustain the growth:} Even under an inverted flow of knowledge ($\alpha_H=0.25<\alpha_A=0.30$) and an inverted learning pattern ($\beta_K=0.03<\beta_A=0.05$), raising the archive gate's accuracy threshold (from $a_0 = 0.5$ to $a_0=0.8$) acts as a “quality firewall”. The stricter admission gate blocks most low-accuracy AI content from entering the long-term corpus, breaking the self-reinforcing collapse loops. As a result, the archive quality $q(t)$ levels off instead of eroding, supervised learning can still push model skill $\theta(t)$ upward, and human expertise $H(t)$ improves rather than decays. Thus, tight gating stabilizes model, humans and corpus, even with inverted flow and learning. 

\vspace{-0.2cm}
\paragraph{(h) Quality dilutes but model improves:}
Despite a strongly inverted archival flow that lets LLMs dominate content creation ($\alpha_H=0.05<\alpha_A=0.5$) and a human preference for AI content over primary study ($\beta_K=0.03<\beta_A=0.05$), the model’s skill steadily rises because the RLHF channel is turned up an order of magnitude higher compared to previous cases ($\eta_{\mathrm{RLHF}}=0.50$ compared to $0.03$ in previous cases).
The high feedback gain of the model means that the high demand for LLM answers, though detrimental to archive quality, helps the model improve heavily. Thus, $\theta(t)$ keeps climbing even as the archive is increasingly injected with mediocre AI text and average quality $q(t)$ drifts downward. Human expertise $H(t)$ plateaus (due to learning from a low-quality archive). Thus, once the model is ``good enough”, treating the archive as a scaffolding for it, and not as the key foundation can prevent model collapse.

\vspace{-0.2cm}
\paragraph{(i) Periodic oscillations:}
Here, the same inverted flow ($\alpha_H=0.05$, $\alpha_A=0.5$) is paired with (i) an extremely soft gate ($\kappa_{\text{gate}}=1.0$, $a_0=0.5$) that admits low-accuracy answers, (ii) a high AI-learning bias ($\beta_A=0.30\gg\beta_K=0.05$), and (iii) a highly sensitive RLHF loop ($\eta_{\mathrm{RLHF}}=0.50$, but with a tiny $Q_{1/2}=100$).
Because query generation becomes steeply reactive to drops in human skill ($K_{1/2} = 200, \kappa_H=0.9$) and task difficulty is set high ($T_\mathrm{difficulty} = 10^4$), even a modest dip in human skill $H(t)$ unleashes a surge of questions. That spike drives up $Q(t)$, which (through RLHF) temporarily boosts $\theta(t)$; the surge of AI answers then swells the archive, dragging $q(t)$ down and, with a short delay, eroding $H(t)$ further.
The coupled overshoot (in $Q(t)$, then $\theta$, then $K(t)$, then $q(t)$, back to $H(t)$) forms an oscillator. Such self-exciting boom-and-bust cycles illustrate how loose content moderation and over-reliance on AI tutoring can destabilize a knowledge ecosystem.

These simulations emphasize the delicate balance needed for sustainable co-evolution of human and AI knowledge. For example, a slight increase in LLM integration can accelerate growth when carefully regulated (as in case~(b),(c)), but mismanagement of archival flow or learning channels can easily lead to stagnation or collapse (cases~(d)–(f)). Our model offers a quantitative foundation for identifying such tipping points and for evaluating both policy-level (e.g.,~content moderation via the admission gate, investment in human learning pathways) and  platform-level levers (e.g.,~proper balance in training modalities) to steer the ecosystem toward long-term quality and resilience.

\vspace{-0.2cm}
\subsection{Real-world Case Studies}
\label{subsec:hybrid_results}

\begin{figure}[]
  \centering
  \begin{subfigure}[b]{0.49\columnwidth} 
    \centering
\includegraphics[width=\textwidth,
                     trim={0.2cm 0.3cm 0.2cm 0.2cm}, clip]{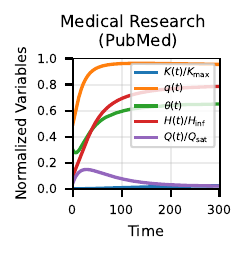}
    \caption{Medical research corpora based on PubMed}
    \label{subfig:Medical}
  \end{subfigure}
    \begin{subfigure}[b]{0.49\columnwidth} 
    \centering
\includegraphics[width=\textwidth,
                     trim={0.2cm 0.3cm 0.2cm 0.2cm}, clip]{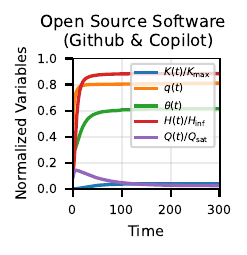}
    \caption{Open source software based on GitHub \& Copilot.}
    \label{subfig:OSS}
  \end{subfigure}\hfill
  \vspace{-0.3cm}
  \caption{Real-world case studies related to two different archives that lead to different growth dynamics.}
  \vspace{-1cm}
\label{fig:hybrid_results}
\end{figure}

In this section, we use platform-level statistics from literature to approximate and compare knowledge archives of two different domains: biomedical research~(PubMed) and open-source software~(GitHub \& Copilot). We show that the model reproduces two different regimes of growth in these platforms and yields insights on sustainable strategies.

\vspace{-0.2cm}
\paragraph{Biomedical research \& clinical decision support (2025).}
PubMed contains tens of millions of indexed citations and has grown approximately 4.5\% annually ($\approx 0.37\%\;\text{mo}^{-1}$) between 2021-2023 on average~\cite{PubMed2025Stats}.\footnote{We have set  $K(0)=100$ as initial condition (see Appendix~\ref{app:numerical_simulation_details}). So this corresponds to an initial absolute growth of $\Delta K \approx 0.0037\times 100 = 0.37$ units per month.}   
Stylometric screenings of biomedical articles suggest that 13.5\% of 2024 submissions were drafted or polished with large language models (LLMs)~\cite{kobak2025delving}. Hence, we map those approximately to $\alpha_H = 0.32\;\text{mo}^{-1}$ ($0.37\times 0.865$) and $\alpha_A = 0.05$ ($0.37\times 0.135$)  in Eq.~\eqref{eq:model_K}. 
Half-life studies of clinical practice guidelines report a median obsolescence of 5.8 years, so we set the knowledge-decay parameter to $\delta_K \simeq 0.010\;\text{mo}^{-1}$ \cite{Shekelle2001Guidelines}.  
A specificity of approximately 99\% is reported in medical diagnostic tasks~\cite{Petmezas2022Ecgreview}, indicating $q_H = 0.99$ while keeping $\delta_q$ small.  
Clinician surveys indicate that decision-support AI must maintain at least 90\% accuracy before they will rely on it, justifying a tight admission gate with $a_0 = 0.90$ and sharpness $\kappa_{\text{gate}} = 20$ \cite{Cabitza2020MinAcc}. 
Finally, more recent LLMs show large performance gains on medical benchmarks~e.g., GPT-4 improves by more than 30 percentage points over GPT-3.5~\cite{Nori2023GPT4Med}. Based on this post-deployment improvement factor, we set $\eta_{\text{RLHF}}=0.30$. 
The resulting configuration in Fig.~\ref{fig:hybrid_results}(a) shows how the corpus quality, and consequently, the model skill have both grown to high levels as a consequence of the increased quality moderation. 

\vspace{-0.2cm}
\paragraph{Open-source software engineering (GitHub + Copilot 2025).}
 GitHub's 2024 Octoverse reports 5.2B human contributions from $\approx100$ million total developers~\cite{Github2024Octoverse} (i.e.,~$\approx 4.33\;\text{mo}^{-1}$ human contributions). Therefore, we set $\alpha_H = 5.0\;\text{mo}^{-1}$.
 GitHub also reports Copilot already generated $\approx46\%$ of developers' code on average across programming languages~\cite{Reddington2023BoostingProductivityGenAI}. 
Hence, we attribute an AI flow of $\alpha_A \approx 4.26\;\text{mo}^{-1}$ (i.e.,~$5.0\;\text{mo}^{-1}\times 0.46/0.54$).  
Knowledge decays faster in software~e.g.,~Stack Overflow answers can become obsolete quickly~\cite{Zhang2019ObsoleteSO}. We model this with an effective knowledge half-life of about 2.5 years, i.e.,
$\delta_K = \ln 2 /(2.5\times 12)\approx 0.023~\mathrm{mo}^{-1}$.  
Experiments find that developers accept roughly one-third of Copilot’s suggestions~\cite{Gao2024}. Based on this statistic, we choose a moderate gate threshold $a_0 = 0.60$ with $\kappa_{\text{gate}} = 8$.
Supervised learning rate is increased ($\eta_{\text{sup}} = 0.12$) and capacity expanded ($K_{\max}=5\times10^{5}$) to reflect the scale of public code corpora.
In this configuration (Fig.~\ref{fig:hybrid_results}(b)),  the archive quality stabilizes at a lower value, though the archive size and human skill improve more (compared to the previous case) due to the increased AI content.

\begin{figure*}[!t]
\centering
\vspace{-0.3cm}
\includegraphics[width=0.5\linewidth]{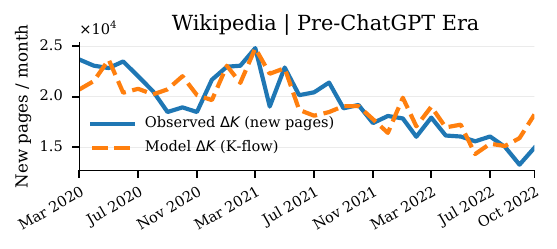}\hfill
\includegraphics[width=0.5\linewidth]{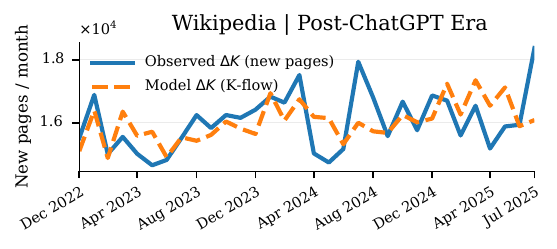}
\vspace{-0.8cm}
\caption{Model calibration with Wikipedia data for pre- and post- ChatGPT eras (33 months each). The monthly flows (solid) vs.\ model-implied flows (dashed) are shown for the flow equation (Eq.~\eqref{eq:model_K}). The estimated parameters are given in Table~\ref{tab:wikipedia-calib}.}
\label{fig:wikipedia-empirical}
\vspace{-0.25cm}
\end{figure*}

Comparison of the two scenarios (medical and open source software corpora) shows how the growth regimes of platforms can be analyzed and compared using the proposed model. Under conservative admission and slower knowledge decay, biomedical corpora converge to high quality and model skill despite relatively modest AI inflow. In contrast, in open source software, the combination of faster obsolescence and a larger AI contribution expands the archive and human skill but stabilizes at a lower quality level. The actionable insights discussed in Sec.~\ref{subsec:numerical_experiments} can be used to optimize the growth of such platforms.

\vspace{-0.2cm}
\subsection{Calibration on Wikipedia for Two Eras}
\label{sec:empirical-wikipedia}

In this section, we estimate the knowledge-flow parameters of Wikipedia using the proposed model \((\alpha_H,\alpha_A,\delta_K\) in Eq.~\eqref{eq:model_K}) for two consecutive, non-overlapping periods of equal length (33 months each), split at the introduction of ChatGPT on Nov.\ 30, 2022:\\
\emph{i.~Pre-ChatGPT Era} (March 2020 to November 2022)\\
\emph{ii.~Post-ChatGPT Era} (December 2022 to August 2025).\\
Post-ChatGPT, the fitted parameters are consistent with a larger AI inflow (\(\alpha_A\) increases) and a lower marginal human inflow (\(\alpha_H\) decreases), filtered by similar gate strictness.

\vspace{0.1cm}
\noindent
\textbf{Calibration process.} For each month \(t\) we use the Wikimedia API to collect: (i) \(\Delta K_t=\) new pages, (ii) \(H_t=\) active editors with \(\geq 5\) edits/month, and (iii) \(Q_t=\) user page-views (measured in millions). The auxiliary components (admission gate \(g(\cdot)\), answer accuracy \(a(\cdot)\), and quality and model dynamics) are fixed (see Appendix~\ref{app:empirical-wiki-robust} for details). This isolates how the human and AI channels map into the archive growth on Wikipedia.

We then estimate the K-flow block \((\alpha_H,\alpha_A,\delta_K)\) by robust least squares with multiple random restarts (16) and box constraints \(\alpha_H\in[10^{-8},10]\), \(\alpha_A\in[10^{-8},5]\), \(\delta_K\in[10^{-6},0.2]\). We integrate Eq.~\eqref{eq:model_K} (for $K(t)$) using the one-step exact update under piecewise-constant monthly controls \(H_t\) and \(Q_t\) (stable for \(\delta_K\ge 0\)) and advance \((q_t,\theta_t)\) with an explicit Euler step.

\vspace{0.1cm}
\noindent
\textbf{Main estimates and fit.}
Table~\ref{tab:wikipedia-calib} summarizes the calibration. In the pre-ChatGPT era, the AI channel in Eq.~\eqref{eq:model_K} is effectively off (\(\alpha_A\approx 10^{-8}\)); growth is explained by the human channel with mild depreciation (\(\delta_K \approx 4.14\times10^{-3}\)). In the post-ChatGPT era, \(\alpha_A\) rises to $0.96$ while $\alpha_H$ drops from $0.7$ to $0.22$, consistent with partial substitution toward AI-mediated additions filtered by the gate.\footnote{Note that $\alpha_H$ is a marginal inflow coefficient, not a content share. The post-ChatGPT era decrease (from $0.70$ to $0.22$) implies a lower per-unit impact of human skill. The realized human contribution depends on $\alpha_H \times H_t$.} Figure~\ref{fig:wikipedia-empirical} shows observed \(\Delta K_t\) versus the calibrated model.

\vspace{0.1cm}
\noindent
\textbf{Mechanisms and robustness checks.}
Beyond the inversion of flow contributions, robustness variants (Appendix~\ref{app:empirical-wiki-robust}) corroborate three mechanisms that also appear in simulations:
\begin{enumerate}[label=(\roman*), leftmargin=10pt, itemsep=2pt, topsep=2pt]
\item \emph{Moderation matters.} Turning the gate off (i.e.,~setting~\(g\equiv 1\)) increases the AI coefficient (e.g., Post-ChatGPT \(\alpha_A\approx 1.10\)) and worsens level-fit (cumulative \(K\)) error.
\item \emph{Demand can lead supply.} Imposing a one-month lag on \(Q\) leaves errors similar and modestly raises \(\alpha_A\).
\item \emph{Small decay.} \(\delta_K\) is small pre-ChatGPT and at the lower bound post-ChatGPT. A joint fit with shared \(\delta_K\) across eras leaves the main insights unchanged (see Appendix~\ref{app:empirical-wiki-robust}), indicating that the shifts in the parameters $\alpha_H, \alpha_A$ are robust to the exact \(\delta_K\) value.
\end{enumerate}

\vspace{0.1cm}
\noindent
\textbf{Interpretation.}
Pre-LLM Wikipedia appears human-led (\(\alpha_A\approx 0\)), with growth tracking human contributions \(H_t\). Post-LLM, the AI content increases while the human channel declines, with gate intensity remaining in a similar range across eras (Appendix~\ref{app:empirical-wiki-robust} provides further robustness tests). These patterns match the theoretical regimes in Sec.~\ref{subsec:numerical_experiments} and illustrate how the proposed model can support platform-level audits to inform policy and design.

\begin{table}[]
\centering
\small
\caption{Wikipedia calibration of \(\alpha_H,\alpha_A,\delta_K\) in Eq.~\eqref{eq:model_K} (archive flow). Additional robustness tests are given in Appendix~\ref{app:empirical-wiki-robust}.}
\vspace{-0.3cm}
\label{tab:wikipedia-calib}
\begin{tabular}{lrrrr}
\toprule
Era (33 mos. each) & \(\alpha_H\) & \(\alpha_A\) & \(\delta_K\) & RMSE\\
\midrule
Pre-ChatGPT  & 0.70081  & \(1.0\times10^{-8}\) & \(4.1416\times10^{-3}\) & 1813.87\\
Post-ChatGPT & 0.224234 & 0.963025             & \(1.00269\times10^{-6}\) & 867.281 \\
\bottomrule
\end{tabular}
\vspace{-0.6cm}
\end{table}

\vspace{-0.3cm}
\section{Conclusion}


This paper offers a systems theoretic view of the modern Web’s human-AI collective knowledge ecosystem. By modeling it as a dynamical system where humans and LLMs co-evolve around a shared archive, we unify three widely discussed risks---\emph{model collapse}, \emph{quality dilution}, and \emph{competence inversion}---within a unified framework. Numerical experiments show that a moderate LLM content generation (compared to human content) coupled with balanced human learning (i.e.,~self-study and LLM queries in a balanced fashion) can sustain healthy growth in archive quality and volume. However, \emph{inverted flow} (LLM-dominated content) and \emph{inverted learning} (over-reliance on LLM answers) create tipping points that trigger collapse of human skill, degrade archives, and eventually weaken LLMs. We identify {policy and design levers} that lead to sustainable growth of the ecosystem: stricter {admission gates on the archive} can prevent most collapse loops; shifting the {LLM training mix} toward human feedback (RLHF) after certain skill threshold can prevent model collapse; and investing in {archive-centric human learning} raises the overall content quality. Calibrating the model on Wikipedia for pre- and post- ChatGPT eras illustrates that the model can be used for platform-level audits in content composition. Our results show that in the post-ChatGPT era,  AI content has accelerated. We also provide two domain-level case studies (medical domain vs.\ open-source software) to illustrate how different growth regimes can emerge in real-world knowledge ecosystems.


\begin{acks}
Authors thank Christoph Riedl (Northeastern University) for feedback during the early phase of this work.
\vspace{-0.2cm}
\end{acks}

\bibliographystyle{ACM-Reference-Format}
\bibliography{WebConf2026_HAICK_Refs}

\appendix

\appendix
\renewcommand\thefigure{S\arabic{figure}}

\vspace{-.cm}
\section*{Appendix}
\section{Limitations and Future Directions}
\label{app:limitations}

Our model is intentionally minimal and mean-field. It aggregates many micro-level processes into five state variables. It does not represent item-level content, user-level heterogeneity, or network structure. It also omits factors such as virality, strategic behavior, platform incentives and norms. Several mechanisms depend on chosen functional forms and parameter values. Different choices could shift regime boundaries and quantitative predictions. 

Our empirical results also have limited scope. Not all state variables are directly observable on most platforms. We therefore use coarse proxies in the case studies (PubMed, Github and Copilot) and in the Wikipedia calibration. Parameter values in the case studies may not be uniquely identifiable from such aggregates. Calibration uncertainty and platform policy changes can also affect the proxies and inferred rates. The Wikipedia calibration targets only the archive-flow sub-system and treats other variables as exogenous. It is a proof-of-concept fit to broad trends and not a complete validation of the coupled system or a causal test. 

Future work should incorporate richer measurements, platform-specific definitions of quality and gating, explicit time lags and bursty dynamics, heterogeneous user groups, and validation across additional domains and intervention settings.

\newpage
\section{Notation and Terminology}
\label{app:terminology}

\begin{table}[htbp]
\centering
\caption{Symbols and parameters used in the five-state Human–AI collective knowledge model}
\vspace{-0.2cm}
\begin{tabular}{ll}
\toprule
\textbf{Symbol} & \textbf{Description} \\
\midrule
\multicolumn{2}{l}{\emph{State variables}}\\
$K(t)$ &  Archive size Eq.~\eqref{eq:model_K} \\
$q(t)$ & Average archive quality \eqref{eq:model_q}\\
$\theta(t)$ & Raw model skill (LLM capacity) Eq.~\eqref{eq:model_theta} \\
$H(t)$ & Average human skill Eq.~\eqref{eq:model_H} \\
$Q(t)$ & Running query volume for LLMs~Eq.~\eqref{eq:model_Q} \\
\addlinespace
\multicolumn{2}{l}{\emph{Auxiliary functions}}\\
$\sigma(\theta)$ & Logistic skill map $\sigma(\theta)=1/(1+e^{-(\theta-\theta_{\text{mid}})})$ \\
$a(\theta,q)$ & Answer accuracy given model skill \& corpus quality \\
$g(a)$ & Smooth admission gate $g(a)=1/(1+e^{-\kappa_{\text{gate}}(a-a_0)})$ \\
$\theta^{*}(K,q)$ & Scaling-law target model skill $\propto q\log(1+K)$ \\
$\mathrm{RLHF}(\theta,Q)$ & LLM learning from human feedback (RLHF)\\
$H_{\max}(K,q)$ & Human capacity ceiling (Hill form, asymptote $H_\infty$) \\
$S(Q)$ & Human learning assistance from LLM assistance \\
$\xi(H)$ & Baseline query generation rate (falls with higher $H$) \\
\addlinespace
\multicolumn{2}{l}{\emph{Archive-growth parameters}}\\
$\alpha_H$ & Human content generation rate (per unit skill) \\
$\alpha_A$ & AI content generation rate per accepted answer \\
$\delta_K$ & Obsolescence / pruning rate of the archive \\
\addlinespace
\multicolumn{2}{l}{\emph{Quality-dynamics parameters}}\\
$q_H$ & Baseline quality of human contributions \\
$\delta_q$ & Spontaneous downward drift of corpus quality \\
\addlinespace
\multicolumn{2}{l}{\emph{Model-learning parameters}}\\
$\eta_{\text{sup}}$ & Supervised scale-up learning rate \\
$\eta_{\text{RLHF}}$ & RLHF boost factor \\
$\theta_{\max}$ & Maximum attainable model skill \\
\addlinespace
\multicolumn{2}{l}{\emph{Human-learning parameters}}\\
$\beta_K$ & Self-study learning rate from the archive \\
$\beta_A$ & Learning rate via AI tutoring \\
$\gamma_H$ & Forgetting / attrition rate \\
$H_\infty$ & Asymptotic human knowledge ceiling \\
$\beta$ & Human learning exponent in the human ceiling\\
$K^\beta_{1/2}$ & Half-saturation constant in the human ceiling \\
$Q_{\text{sat}}$ & Saturation threshold for AI tutoring returns \\
\addlinespace
\multicolumn{2}{l}{\emph{Admission-gate parameters}}\\
$a_0$ & Accuracy threshold that AI answers must exceed \\
$\kappa_{\text{gate}}$ & Steepness of logistic gate $g(a)$ \\
\addlinespace
\multicolumn{2}{l}{\emph{Query-flow parameters}}\\
$\rho_Q$ & Query resolution / dissipation rate \\
$\kappa_H$ & Sensitivity of query generation to human skill \\
$T_\text{difficulty}$ & Task difficulty \\
\addlinespace
\multicolumn{2}{l}{\emph{Scaling-law / RLHF constants}}\\
$K_{\max}$ & Reference corpus size in the scaling law \\
$Q_{1/2}$ & Half-saturation constant in the RLHF term \\
\bottomrule
\end{tabular}
\label{tab:terminology}
\end{table}

\section{Well-posedness and forward invariance of the Model}
\label{app:wellposedness}

Let $x=(K,q,\theta,H,Q)$ evolve under Eq.~\eqref{eq:model} with the domain
\[
\mathcal D \;=\; \{\,K\!\ge 0,\; q\!\in[0,1],\; \theta\!\in[0,\theta_{\max}],\; H\!\ge 0,\; Q\!\ge 0\,\}.
\]

\smallskip
\noindent\textbf{Well-posedness (existence and uniqueness).}
For any initial $x(0)\in\mathcal D$ with $K(0)>0$, the vector field in Eq.~(1) is locally Lipschitz on a neighborhood of the trajectory, so the system admits a unique solution $x(t)$ for all $t\ge 0$.
Intuitively, the only potentially delicate term is the $K^{-1}$ factor in $\dot q$;
because $K(t)$ remains strictly positive once $K(0)>0$ (the archive has nonnegative inflows and linear decay),
these coefficients remain bounded, ensuring standard ODE well-posedness.

\smallskip
\noindent\textbf{Forward invariance of the physical domain.}
The faces of $\mathcal D$ are invariant under Eq.~(1):
$K$ has only nonnegative inflows (human $+\,$gated-AI) and linear decay, so $K(t)\ge 0$;
$q\in[0,1]$ is driven toward a convex combination of $q_H\in[0,1]$ and $a(\theta,q)\in[0,1]$ with nonnegative weights, minus $\delta_q q$,
which keeps $q$ in $[0,1]$; $\theta$ is pulled toward $\theta^\star\in[0,\theta_{\max}]$ with an RLHF term bounded by $(\theta_{\max}-\theta)$, which keeps
$\theta\in[0,\theta_{\max}]$; $H$ and $Q$ have nonnegative inflow terms and linear dissipation, so $H,Q\ge 0$.
Consequently, trajectories starting in $\mathcal D$ remain in $\mathcal D$ for all $t\ge 0$.

\smallskip
\noindent\textbf{Global boundedness.}
Under the model’s primitives,
the drivers $(q,\theta,H,Q)$ stay uniformly bounded; hence the net inflow
$A(t)=\alpha_H H(t)+\alpha_A Q(t) g(a(t))$ is bounded, and $K$ remains bounded as well.
This guarantees numerically stable simulations and parameter fitting.

\smallskip
\noindent\textbf{Discrete-time implementation.}
For monthly calibration we advance $K$ with the exact one-step solution of
$\dot K=A_t-\delta_K K$ under piecewise-constant inputs (Appendix~C.3)
and step $(q,\theta)$ once per month (clipping $q$ to $[0,1]$).
These updates preserve the forward-invariance of $\mathcal D$ at the resolution used in our experiments.

\section{Details on Numerical Simulations}
\label{app:numerical_simulation_details}

\begin{table*}
\setlength{\tabcolsep}{5pt}
\renewcommand{\arraystretch}{1.15}
\caption{Full parameter configurations corresponding to the nine numerical experiments in Fig.~\ref{fig:numerical_results} in Sec.~\ref{subsec:numerical_experiments} scenarios. Boldface marks parameters that differ from the case~(b)~Healthy Growth baseline. In all nine cases, the fixed parameters are:  $\delta_{K}=0.01$, $q_{H}=0.95$, $\delta_{q}=0.001$, $\theta_{\mathrm{mid}}=0$, $\theta_{\max}=1.0$, $K_{\max}=10^{4}$, $\gamma_{H}=0.05$, $Q_{\mathrm{sat}}=500.0$, $H_{\infty}=100$, and $\rho_{Q}=0.01$.
 }
 \label{tab:params_numerical_experiments}
\begin{tabular}{lccccccccc}
\toprule
 & \textbf{(a)} & \textbf{(b)} & \textbf{(c)} & \textbf{(d)} & \textbf{(e)} & \textbf{(f)} & \textbf{(g)} & \textbf{(h)} & \textbf{(i)} \\
\textbf{} &
\textit{No LLMs} &
\textit{Healthy} &
\textit{Healthy Growth} &
\textit{Inv.\ Flow} &
\textit{Inv.\ Learn.} &
\textit{Inv. Flow \& } &
\textit{Gate Tightening} &
\textit{Model Recovery} &
\textit{Oscill.} \\
\textbf{Param.} &
\textit{} &
\textit{Growth} &
\textit{\& Tighter Gate} &
\textit{} &
\textit{} &
\textit{Inv. Learn} &
\textit{Stops Collapse} &
\textit{due to RLHF} &
\textit{} \\
\midrule
\textbf{$\alpha_H$}      & 0.5 & 0.5 & 0.5 & \textbf{0.05} & 0.5 & \textbf{0.05} & \textbf{0.25} & \textbf{0.05} & \textbf{0.05} \\
\textbf{$\alpha_A$}      & \textbf{0.00} & 0.05 & 0.05 & \textbf{0.5} & 0.05 & \textbf{0.5} & \textbf{0.3} & \textbf{0.5} & \textbf{0.5} \\
\textbf{$\beta_K$}       & 0.05 & 0.05 & 0.05 & 0.05 & \textbf{0.03} & \textbf{0.03} & \textbf{0.03} & \textbf{0.03} & 0.05 \\
\textbf{$\beta_A$}       & \textbf{0.00} & 0.03 & 0.03 & 0.03 & \textbf{0.05} & \textbf{0.05} & \textbf{0.05} & \textbf{0.05} & \textbf{0.3} \\
\textbf{$\eta_{\mathrm{sup}}$}  & \textbf{0.00} & 0.05 & 0.05 & 0.05 & 0.05 & 0.05 & 0.05 & 0.05 & 0.05 \\
\textbf{$\eta_{\mathrm{RLHF}}$} & \textbf{0.00} & 0.03 & 0.03 & 0.03 & 0.03 & 0.03 & 0.03 & \textbf{0.5} & \textbf{0.5} \\
\textbf{$a_0$}           & \textbf{0.0} & 0.5 & \textbf{0.8} & 0.5 & 0.5 & 0.5 & \textbf{0.8} & 0.5 & 0.5 \\
\midrule
$Q_{1/2}$                & 5000.0 & 5000.0 & 5000.0 & 5000.0 & 5000.0 & 5000.0 & 5000.0 & 5000.0 & \textbf{100.0} \\
$\beta$ & 0.9 & 0.9 & 0.9 & 0.9 & 0.9 & 0.9 & 0.9 & 0.9 & \textbf{4} \\
$K_{1/2}$                & 300 & 300 & 300 & 300 & 300 & 300 & 300 & 300 & \textbf{200} \\
$\xi_0$                  & \textbf{0.0} & 2.0 & 2.0 & 2.0 & 2.0 & 2.0 & 2.0 & 2.0 & 2.0 \\
$\kappa_H$               & 0.05 & 0.05 & 0.05 & 0.05 & 0.05 & 0.05 & 0.05 & 0.05 & \textbf{0.9} \\
$T_\mathrm{difficulty}$    & 10.0 & 10.0 & 10.0 & 10.0 & 10.0 & 10.0 & 10.0 & 10.0 & \textbf{10000.0} \\
$\kappa_{\mathrm{gate}}$ & 10.0 & 10.0 & 10.0 & 10.0 & 10.0 & 10.0 & 10.0 & 10.0 & \textbf{1.0} \\
\bottomrule
\end{tabular}
\end{table*}

This section provides details on the numerical experiments given in Sec.~\ref{subsec:numerical_experiments} and the two domain case studies given in Sec.~\ref{subsec:hybrid_results}. These details and the provided code help reproduce all results. 

\paragraph{Integrator and grid.}
We integrate the continuous-time system with \texttt{scipy.integrate.solve\_ivp} using \texttt{method="RK45"} (Dormand–Prince, adaptive step). We integrate the ODEs with an adaptive RK45 solver using a relative tolerance of $10^{-6}$ and an absolute tolerance of $10^{-9}$. The simulation runs on $t\in[0,T]$, and we request outputs on a uniform grid of $T{+}1$ points (i.e., at every integer time $0,1,\dots,T$).
We use $T=1000$ for regime plots and extend to longer horizons (e.g., $T=5000$) only when transients are slow.

\paragraph{Scenarios in Sec.~\ref{subsec:numerical_experiments}.}
Each subplot in Fig.~\ref{fig:numerical_results} is based on the parameters specified in Table~\ref{tab:params_numerical_experiments}. The states variables are initialized at
\[
y_0=[K_0,q_0,\theta_0,H_0,Q_0]=[100.0,\;0.5,\;0.3,\;10.0,\;50.0].
\]

\paragraph{Case studies in Sec.~\ref{subsec:hybrid_results}.}
The \emph{Medical Research} and \emph{Open Source Software} panels reuse the same integrator, tolerances, grid, and $y_0$, differing only in their parameter configurations given in Table~\ref{tab:params_hybrid_experiments}.

\paragraph{State normalization.}
After integration, we plot the five state variables (normalized when relevant): $K(t)/K_{\max}$, $q(t)$, $\theta(t)$, $H(t)/H_\infty$, and $Q(t)/Q_{\text{sat}}$.

\paragraph{Reproducibility details.} All codes for the above simulation pipleline are publicly available.

\FloatBarrier 
\begin{table}[!htbp]
\caption{Full parameter configurations corresponding to the two case studies in Fig.~\ref{fig:hybrid_results} in Sec.~\ref{subsec:hybrid_results}.}
\label{tab:params_hybrid_experiments}
\begin{tabular}{lcc}
\toprule
\textbf{Parameter} & \textbf{Medical Research} & \textbf{Open Source Software} \\
\midrule
$\alpha_H$            & 0.32  & 5.0 \\
$\alpha_A$            & 0.05  & 4.26 \\
$\delta_{K}$          & 0.010 & 0.023 \\
$q_{H}$               & 0.99 & 0.90 \\
$\delta_{q}$          & 0.0005 & 0.002 \\
\midrule
$\theta_{\mathrm{mid}}$ & 0   & 0 \\
$\theta_{\max}$         & 1.0 & 1.0 \\
$K_{\max}$              & $10^{5}$ & $5\times 10^{5}$ \\
$\eta_{\mathrm{sup}}$   & 0.10 & 0.12 \\
$\eta_{\mathrm{RLHF}}$  & 0.30 & 0.25 \\
$Q_{1/2}$               & 8000.0 & 5000.0 \\
\midrule
$\beta_{K}$            & 0.20 & 0.30 \\
$\beta_{A}$            & 0.05 & 0.10 \\
$\gamma_{H}$           & 0.02 & 0.03 \\
$Q_{\mathrm{sat}}$     & 1000.0 & 500.0 \\
$H_{\infty}$           & 100 & 100 \\
$K_{1/2}$              & 300 & 300 \\
$\beta$ & 0.9 & 0.9 \\
\midrule
$\xi_{0}$              & 2.0 & 2.0 \\
$\kappa_{H}$           & 0.05 & 0.05 \\
$\rho_{Q}$             & 0.02 & 0.02 \\
$T_\mathrm{difficulty}$  & 10.0 & 10.0 \\
\midrule
$a_{0}$                & 0.90 & 0.60 \\
$\kappa_{\mathrm{gate}}$ & 20.0 & 8.0 \\
\bottomrule
\end{tabular}
\end{table}

\section{Details on Wikipedia Experiment}
\label{app:empirical-wiki-robust}

This section details the end-to-end pipeline used to calibrate the Wikipedia knowledge-flow block (using Eq.~\eqref{eq:model_K}).

\vspace{0.25em}
\paragraph{E.1 Data sources and preprocessing.}
We use Wikimedia REST AQS endpoints at monthly granularity for \texttt{en.wikipedia}:
\begin{itemize}
  \item \textbf{New pages:} target series $\Delta K_t$ (flow). We request content pages (if unavailable, we choose all page types).
  \item \textbf{Active editors: } human-skill proxy $H_t$. We sum buckets 5--24, 25--99, and 100+ edits. If any bucket is absent, we use those with at least one edit.
  \item \textbf{Pageviews: } demand proxy $Q_t$, rescaled to millions: $Q^{(M)}_t = Q_t/10^6$.
\end{itemize}
\textbf{Time bounds.} Two equal-length eras are used in the paper:
\emph{Pre-ChatGPT} = 2020-03-01 to 2022-11-30; \emph{post-ChatGPT} = 2022-12-01 to 2025-08-30. Each month is aligned to month-end; series are inner-joined on month.

\vspace{0.1cm}
\noindent
\textbf{Reasoning behind the chosen proxies:}\\
\emph{(i) New pages $\Delta K_t$.} We use Wikimedia “new pages” as the flow into the archive because it is a direct, platform-native measure of stock expansion; restricting to \texttt{page\_type=content} prioritizes encyclopedic content and avoids effects from talk/user pages. This aligns with the $K$-equation, which models net additions minus depreciation.

\emph{(ii) Active editors $H_t$.} Summing the 5–24, 25–99, and 100+ edit buckets captures the mass of sustained contributors (beyond one-off edits) and thus serves as a practical proxy for aggregate human skill/effort available to contribute in a given month. Where bucket availability varies, falling back to “at least 1 edit” category preserves coverage without biasing trends.

\emph{(iii) Pageviews $Q_t$.} Monthly user pageviews (all-access, \texttt{agent=user}) are a transparent proxy for information-seeking demand that co-moves with LLM query pressure in the coupled system; the $K$-flow block only requires a demand signal that scales gated AI inflow, not exact API call counts. Using user-only traffic reduces bot contamination, and the robustness checks (lagging $Q$, gate variants, $Q_{1/2}$ sensitivity) indicate conclusions are not sensitive to this particular operationalization.

\vspace{0.25em}
\paragraph{E.2 Fixed hyperparameters.}
We construct the initial stock for each era by summing monthly new pages from 2001-01 through the month immediately preceding the era start: $K_0 = \sum_{\tau=\text{2001-01}}^{t_0-1}\Delta K_\tau$ where $t_0$ is the era start. A single $K_{\max}$ is used for both eras and set to $1.25\times$ the cumulative new pages from 2001-01 through 2025-08. The auxiliary (fixed) parameters are:
\[
\begin{gathered}
q_H=0.85,\quad \delta_q=5{\times}10^{-4},\quad
\theta_{\max}=1,\quad \theta_{\text{mid}}=0, \quad
\eta_{\text{sup}}=0.02,\\\eta_{\text{RLHF}}=0.05,\quad
Q_{1/2}=5000,\quad a_0=0.60,\quad \kappa_{\text{gate}}=10.
\end{gathered}
\]

\vspace{0.25em}
\paragraph{E.3 Estimation.}
We estimate the knowledge-flow parameter triplet $\Phi_K=(\alpha_H,\alpha_A,\delta_K)$ \emph{separately} for each era via robust nonlinear least squares against the flow target $\Delta K_t$. Let
\[
A_t=\alpha_H H_t+\alpha_A\,Q^{(M)}_t\,g(a_t).
\]
The state variable $K$ uses the exact one-step update under piecewise-constant inputs within month:
\[
K_{t+1}=
\begin{cases}
\bigl(K_t - A_t/\delta_K\bigr)\,\mathrm{e}^{-\delta_K} + A_t/\delta_K, & \delta_K>0,\\[4pt]
K_t + A_t, & \delta_K=0.
\end{cases}
\]
The state variables $(q_t,\theta_t)$ advance by explicit Euler using the fixed dynamics in Sec.~\ref{subsec:model}; $q_t$ is clipped to $[0,1]$ each step. We minimize a median-absolute-deviation (MAD) scaled residual with a soft-$\ell_1$ loss over $16$ random restarts. Parameter bounds are: $(\alpha_H,\alpha_A,\delta_K)\in[10^{-8},10]\times[10^{-8},5]\times[10^{-6},0.2]$. Initial $(q_0,\theta_0)$ are $(0.85,0.3)$. We also report level ($K$) errors implied by the flow ($\dot{K}$) fit, and we optionally re-fit directly to levels ($K$) for a consistency check.

\vspace{0.25em}
\paragraph{E.4 Baseline calibration.}
Using the flow objective and the fixed block above the estimates shown in Table~\ref{tab:wikipedia-calib} and the comparison between observed and fitted flows is shown in Fig.~\ref{fig:wikipedia-empirical}.

\vspace{0.25em}
\paragraph{E.5 Robustness variants.}
We vary the fit target, the gate, the demand timing, and key fixed hyperparameters. Each variant reuses the same data.
\begin{itemize}
  \item \textbf{Fit target:} {flow} (baseline) vs.\ {level}. Under {level}, post-ChatGPT era remains AI-on with $\alpha_A\approx 1.0194$ and $\alpha_H\approx 0.2225$, and achieves a lower $K$-level RMSE ($\approx 1304.75$) than the level series implied by the flow objective—consistent with mild model mismatch noise.
  \item \textbf{Gate off:} $g(a)\equiv 1$. Pushes {post-ChatGPT} $\alpha_A$ up ($\approx 1.0970$) and worsens level RMSE to $\sim 3199.6$, underscoring moderation’s role.
  \item \textbf{Demand lag:} replace $Q^{(M)}_{t}$ by $Q^{(M)}_{t-1}$. Errors are similar; {post-ChatGPT} $\alpha_A$ shifts modestly ($\approx 0.8239$ under {level} fit).
  \item \textbf{$K_{\max}$ sensitivity:} multiplier $1.25$ (baseline) vs.\ $1.50$. Conclusions unchanged (post-ChatGPT era $\alpha_A\approx 0.9518$).
  \item \textbf{$Q_{1/2}$ sensitivity:} $Q_{1/2}=5000$ (baseline) vs.\ $2500$. Small parameter drift; qualitative pattern unchanged.
  \item \textbf{Joint $\delta_K$:} a shared $\delta_K\approx 2.45\times 10^{-5}$ across eras leaves post-ChatGPT parameters close to baseline; no qualitative change. 
\end{itemize}

\vspace{0.5cm}
\noindent
{\bf Code Availability and Reproducibility: } Codes related to numerical and empirical results are publicly available on GitHub:
\url{https://github.com/EnExEm/HAICK}

\end{document}